\documentclass[11pt]{article}
\usepackage{coling2020}
\usepackage{times}
\usepackage{url}
\usepackage{latexsym}
\usepackage{graphicx}
\usepackage{amsmath}
\usepackage{bm}
\usepackage{amssymb}
\usepackage{algorithm}
\usepackage{subcaption}
\usepackage{wrapfig}
\usepackage{booktabs}
\usepackage{algpseudocode}
\usepackage{siunitx}
\usepackage{gensymb}
\usepackage[title]{appendix}

\usepackage{tikz}

\graphicspath{{./imgs/}}
\DeclareMathOperator*{\argmax}{arg\,max}

\colingfinalcopy %

\title{Autoregressive Reasoning over Chains of Facts with Transformers}

\author{Ruben Cartuyvels, \, Graham Spinks, \, Marie-Francine Moens \\
  LIIR lab, KU Leuven, Belgium \\
  {\tt \{first\}.\{last\}@kuleuven.be, \, sien.moens@kuleuven.be}
}
\date{}

\begin{document}
\maketitle

\begin{abstract}
    This paper proposes an iterative inference algorithm for multi-hop explanation regeneration, that retrieves relevant factual evidence in the form of text snippets, given a natural language question and its answer.
    Combining multiple sources of evidence or facts for multi-hop reasoning becomes increasingly hard when the number of sources needed to make an inference grows.
    Our algorithm copes with this by decomposing the selection of facts from a corpus autoregressively, conditioning the next iteration on previously selected facts.
    This allows us to use a pairwise learning-to-rank loss. 
    We validate our method on datasets of the TextGraphs 2019 and 2020 Shared Tasks for explanation regeneration.
    Existing work on this task either evaluates facts in isolation or artificially limits the possible chains of facts, thus limiting multi-hop inference. 
    We demonstrate that our algorithm, when used with a pre-trained transformer model, outperforms the previous state-of-the-art in terms of precision, training time and inference efficiency.
\end{abstract}

\blfootnote{
This work is licensed under a Creative Commons Attribution 4.0 International License. \\
License details: \url{http://creativecommons.org/licenses/by/4.0/}.
}

\section{Introduction} \label{sec:intro}

The task of multi-hop explanation generation 
has recently received interest as it could be a stepping-stone towards general multi-hop inference over language. 
Multi-hop reasoning requires algorithms to combine multiple sources of evidence.
This becomes increasingly hard when the number of required facts for an inference grows, because of the exploding number of combinations and phenomena such as semantic drift \cite{fried2015higher,jansen2018multi}.
The WorldTree dataset was designed specifically for ($>$2)-fact inference \cite{jansen2018worldtreev1,xie2020worldtreev2}: it consists of elementary science exam questions that can be explained by an average of 6 facts from a complementary dataset of textual facts.
The explanation regeneration task as in the TextGraphs Shared Tasks \cite{jansen2019textgraphs,jansen2020textgraphs} 
asks participants to retrieve and rank relevant facts (given one of these natural language questions and its answer as input\footnote{
Systems that do not require the answer, but that retrieve facts based on a question only, could be of great utility: the retrieved facts could be used to infer the answer to the question. However in this paper we follow the task definition of \cite{jansen2019textgraphs}, who define explanation regeneration as a stepping-stone task for multi-hop inference. The method we propose is equally applicable when the answer is not given, but evaluating this setting is left for future work.
}) such that the top-ranked facts explain the answer to the question.
An example is shown in the upper left part of fig. \ref{fig:overview} (and more in appendix \ref{app:data-examples}).

As each question-answer pair potentially has many supporting facts, %
evaluating all 
combinations of facts is computationally prohibitive. 
Previous work remedies this by computing scores for facts in isolation, %
or by severely limiting the number of combinations of facts %
\cite{das2019chains,banerjee2019asu,chia2019red}. The latter is done by considering combinations of few facts only and/or by reranking combinations of only the top retrieved facts by a simpler method. Both approaches limit multi-hop reasoning as facts are not combined
or too many facts are ignored by the simple first-stage retriever.

In this paper we propose a method to retrieve facts that does build long chains of facts, while being efficient. %
During training, a pre-trained neural language model encodes the question-answer pair as well as a randomly selected combination of ground-truth facts, before evaluating candidate facts. For efficiency, only a set of neighborhood facts (which we call `visible' facts) is considered. This set is obtained 
by selecting the nearest facts by tf-idf distance to the question, answer and set of ground truth facts.
During inference, 
facts are ranked iteratively: in each iteration, the highest scoring fact is selected and encoded together with previously selected facts and the question-answer pair, so the next ranking is always conditioned on the current set of chosen facts.
At each inference step, the set of visible facts consists of
the nearest facts to the already selected facts and question-answer pair. 
This autoregressive formulation 
and the definition of neighborhoods together enable the use of losses that incorporate 
interactions between different facts, like the pairwise RankNet loss from the learning-to-rank literature \cite{burges2005ranknet}.

Most methods in earlier work use some form of rank-rerank set-up, but none dynamically expand the set of reranked facts, and the reranked set is usually too small \cite{das2019chains,chia2019red,banerjee2019asu}. 
\newcite{banerjee2019asu} and \newcite{chia2019red} use iterative schemes, but the former only reconsiders the top 15 initially retrieved facts 
and the latter uses a term frequency based approach.
\newcite{das2019chains} consider chains of facts during training and inference, but only up to length 2, because of the explosive number of combinations.
All of the above systems use pointwise losses only.
While components of our method are inspired by previous work, we are the first to put them together in a principled way, so that they enable each other.
Our neighborhoods enable both our iterative inference procedure to efficiently build chains of up to 10 facts, and the use of more informative losses. Our training is designed to support iterative inference and to close the gap between training and inference as much as possible.

\textbf{Contributions.} (1) By defining dynamically growing neighborhoods of facts for our model to operate in, we limit computational cost without severely limiting the range of our method,
(2) We define a new autoregressive training and inference method to evaluate facts within the context of other facts; 
(3) We apply a learning-to-rank loss that successfully exploits interactions between facts, leading to an improvement in MAP score over previous baselines.

\section{Related Work} \label{sec:related}

Explanation regeneration was promoted as a TextGraphs 2019 Shared Task \cite{jansen2019textgraphs}. We briefly summarize systems proposed in 2019. Most methods finetuned a pre-trained transformer like BERT \cite{devlin2019bert} in a learning-to-rank set-up, as reranker with the question and answer as query and facts as documents, like %
\newcite{nogueira2019passage}.
\newcite{das2019chains} use BERT to classify chains of 2 facts, drawn from initially retrieved facts by a tf-idf retriever and facts with words in common with those facts.
Their idea is similar to ours, but due to the high number of combinations, they are limited to explanations of only 2 facts. 
As a second approach, they use BERT like %
\newcite{nogueira2019passage} to rank single facts: but Das et al. simply rank all facts instead of the top-$T$ initially retrieved ones, which is computationally expensive for the larger 2020 dataset.

\newcite{chia2019red} use an iterative scheme where the tf-idf representation of the question is aggregated with the tf-idf representations of previously selected facts for retrieval.
They compare this to a BERT based approach like described above. 
In contrast, selected facts and the question in our method are encoded as one paragraph with BERT in a trainable way, which allows for more complex relations to be learnt.
\newcite{banerjee2019asu} adds gold facts as context during training and scores facts individually with BERT during inference. The top facts are reranked by iterative rescoring, which only uses the originally computed scores and precomputed (hence untrainable) sentence embeddings.
\newcite{dsouza2019svmrank} propose a pairwise learning-to-rank approach with SVMs. 

\newcite{das2018retriever} propose a retriever-reader system that iteratively retrieves and `reads' relevant paragraphs from a large text corpus for open-domain QA. High level analogies can be drawn between their system and ours, but their architecture involves separate query and document encoders, a recurrent reasoner and a specialized reading comprehension model. Their system is trained end-to-end for QA.

In conclusion, our work has similarity to the widely used retrieval-reranking paradigm, 
but the initially retrieved set of facts is dynamically extended based on selected facts. Our training and inference method to evaluate facts within the context of other facts successfully models interactions between different facts, framing the task more as a reading comprehension task.

\section{Proposed Approach}  \label{sec:proposed}

\begin{figure}
	\centering %
	\includegraphics[trim={0cm 0cm 0cm 0cm},clip,width=0.99\textwidth]{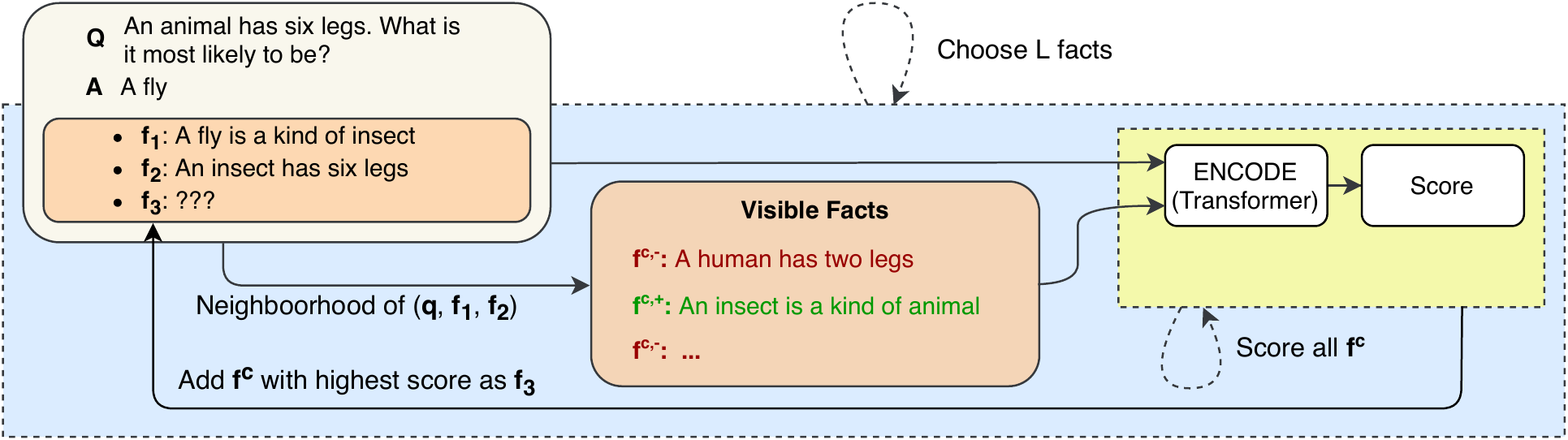}
	\caption{An overview of ARCF during inference. The computed score represents $P(\bm{f}_{3} \mid \bm{f}_{1,2}, \bm{q})$.
	}
	\label{fig:overview}
\end{figure}

This section describes our proposed method, which we call `Autoregressive Reasoning over Chains of Facts' (ARCF). ARCF consists of an initial retrieval component described in §\ref{ssec:fact-lands}, a learning-to-rank training scheme and an iterative inference procedure (both in §\ref{sec:rankingCFs}, fig. \ref{fig:overview} shows the latter schematically).

\subsection{Task Description} \label{ssec:task}

Given is a dataset $\mathcal{D}_{f}$ of facts in text form (size $D_f$). Each fact $\bm{f}$ consists of word tokens: 
$\bm{f} = [w_1, w_2, ..., w_{Z_f}]$. 
Further given is a training, validation and test dataset $\mathcal{D}_{qa}$ containing multiple choice questions (size $D_{qa}$). Each question is concatenated with answer options:
4 (or so) multiple choice answers, with the correct one marked.
Question and answer(s) together form a query $\bm{q} = [w_1, w_2, ..., w_{Z_q}]$.   %
All queries in $\mathcal{D}_{qa}$ can be explained by 1 to 22 gold facts in $\mathcal{D}_{f}$. 
Gold facts for a question are marked with `grounding', `central' or `lexical glue', depending on their role in explaining the question. Central facts are key in explaining the question, lexical glue links facts together (e.g., by synonymy or taxonomy relations) and grounding facts connect facts to the question\footnote{
We refer the reader to \cite{jansen2018worldtreev1,xie2020worldtreev2} for more information.
}.

The task for a given $\bm{q} \in \mathcal{D}_{qa}$ is to rank all facts in $\mathcal{D}_{f}$, with gold facts ranked higher than irrelevant facts. 
The \textit{answer} in this context is the answer to a question, always encoded together with the question in $\bm{q}$, and not the output target of the task. When we write `query', we mean an instance of $\bm{q}$, i.e., a question concatenated with its answer. The output target is the set of gold facts $\bm{f}^*_{1,...,G}$ for a $\bm{q}$.
The mean average precision (MAP)
of the gold facts in the computed ranking is calculated as evaluation metric.
We use the notation $\bm{f}_{1,...,N}$ for an intermediate set of facts, $\bm{f}_{1,...,M}$ for a completed set and $\bm{f}^*_{1,...,G}$ for the gold set for a $\bm{q}$, with $G$ the number of gold facts. We call a concatenation of a query with a number of facts a prefix: $\bm{p} = [\bm{q}\mid\bm{f}_{1,...,N}]$. Appendix \ref{app:data-examples} shows example $\bm{q}$'s and gold facts.

\subsection{Model}

We use a neural language encoder to compute a function $f_\theta: \mathcal{V}^Z \mapsto \mathbb R$. The input is the concatenation of a query and a number of facts $[\bm{q} \mid \bm{f}_{1,...,N} ]$. The output is a scalar score $s$, indicating 
how well the last of the concatenated facts (the candidate fact) fits in the explanation for the query. The process is iterated: a chosen fact is appended to $\bm{f}_{1,...,N}$ and a new score is computed.
We use pre-trained transformer models, like BERT or RoBERTa \cite{devlin2019bert,liu2019roberta,vaswani2017attention}, 
because the task dataset is relatively small, and this allows for the reasoning to incorporate external knowledge. 

\subsection{Fact - Question Neighborhoods} \label{ssec:fact-lands}

\paragraph{Neighborhoods.} The first step of the method 
consists of computing neighborhoods of visible facts, 
for each question and corresponding set of facts,
denoted by $\textit{vis}_k: \mathcal{D}_{f} \cup \mathcal{D}_{qa} \mapsto \mathcal{D}_{f}^K$. To retain tractability, facts from neighborhoods $\textit{vis}_k(\cdot)$ will later be ranked, while facts outside $\textit{vis}_k(\cdot)$ are out of consideration. In contrast to classic rank--(neural) rerank approaches, our neighborhood (initially retrieved set that will be reranked) will expand dynamically.
Pairwise distances between all facts and between questions and facts are precomputed. A fact $\bm{f}^c$ is visible from $[\bm{q} \mid \bm{f}_{1,...,N} ]$ if it is one of the $k$ nearest facts either to $\bm{q}$, or to one of the $\bm{f}_{1,...,N}$ (denoted as $\textit{nearest}_k$). We use $K_N \leq (N+1)\cdot k$ to refer to the cardinality\footnote{
$N+1$ for $N$ facts and $1$ query, $\leq$ because the neighborhoods might overlap. 
} of $\textit{vis}_k(\cdot)$. Formally: $\textit{vis}_k(\bm{q},\bm{f}_{1,...,N}) = \bigcup_{\bm{x} \in \{\bm{q},\bm{f}_{1,...,N}\}} \textit{nearest}_k(\bm{x})$.

For a given question and (possibly empty) set of facts, the algorithm should be able to retrieve the set of visible facts, but is agnostic to how this set or the distances are computed. %
Given the limited size of $\mathcal{D}_{f}$ and $\mathcal{D}_{qa}$, if we use an inexpensive distance metric, the overhead of computing all pairwise distances is small. For larger datasets or for distances that are expensive to compute, the overhead might not be negligible. Approximate nearest-neighbor methods could then be used.

\begin{wraptable}{l}{34mm} %
    \fontsize{9}{11}\selectfont
	\centering
        \begin{tabular}{r c } 
		$k$ & Fraction \\ \toprule
		90 & 0.90 \\
		130 & 0.95 \\
		180 & 0.97 \\
		290 & 0.99 \\
	\end{tabular}
	\caption{Mean fraction of gold facts reachable.} \label{tab:nearest-k}
\end{wraptable}

\paragraph{Distances.} We tried computing the distances as distance between \textit{tf-idf} vectors, as Word Mover's Distance \cite{kusner2015word}, as the distance between sentence embeddings computed by a pre-trained BERT or SentenceBERT \cite{reimers2019sentencebert}, and as the reciprocal number of words in common (lexical overlap). 
We compared the distance metrics by the fraction of gold facts, for a given $k$, that could be reached in an unlimited number of `hops'\footnote{
A hop is defined as an imagined `move' from either a $\bm{q}$ or $\bm{f}$ to another visible fact.
} via \textit{gold} facts from $\textit{vis}_k(\cdot)$, starting from a $\bm{q} \in \mathcal{D}_{qa}$. 
We only computed the fractions on the training data to prevent test set leakage.
We found {tf-idf} to work best for all $k$: table \ref{tab:nearest-k} shows some indicative ratios (mean over $\bm{q} \in \mathcal{D}_{qa}$). The fact that {tf-idf} works best here %
can be explained by the fact that the dataset is well curated and terminology is uniform across facts.

\newcite{das2019chains} 
construct a connectivity graph between facts, which they use to extend the set of initially retrieved facts by tf-idf.
They use lexical overlap as criterion for being connected,
resulting in a dense graph and leading to an explosive number of chains.
In contrast, parameter $k$ in our definition of $\textit{vis}_k$ allows for easy and precise control of the size of neighborhoods.

\subsection{Autoregressive Ranking of Candidate Facts} \label{sec:rankingCFs}

We propose to decompose the conditional probability distribution over rankings of facts autoregressively: %
\begin{equation} \label{eq:autoregr}
	P(\bm{f}_{1,...,M} \mid \bm{q}) \, = \, \prod_{i=1}^{M} \, P(\bm{f}_{i} \mid \bm{f}_{1,...,i-1}, \bm{q}) ,
\end{equation}
where $\bm{f}_1$ is the highest ranked fact. At each iteration, we compute $P(\bm{f}_{i} \mid \bm{f}_{1,...,i-1}, \bm{q})$ for all visible candidate facts $\bm{f}_{i}$, and we select the fact with the highest probability to be ranked at position $i$. The unnormalized probability for $P(\bm{f}_{i} \mid \bm{f}_{1,...,i-1}, \bm{q})$ is computed by our scoring function as $s = f_\theta([\bm{q} \mid \bm{f}_{1,...,i-1} \mid \bm{f}_i])$.
For the task at hand, only the interclass order in the produced ranking is relevant: relevant facts should be ranked higher than irrelevant facts. The intraclass order, i.e., the relative order of relevant facts and of irrelevant facts, does not affect the MAP (§\ref{ssec:task}). 
Eq. \ref{eq:autoregr} can thus be understood as the decomposition of selecting a set of facts jointly into selecting one fact at a time (conditioned on previously selected facts).
Scoring all combinations of facts is definitely infeasible, while scoring facts independently is too simplifying.
The decomposition aims to strike a balance between the two.

Conditioning the selection of facts on previously selected facts brings several advantages (compared to scoring facts independently as %
\newcite{nogueira2019passage}). First, it enables the incremental building of chains of reasoning.
The role of many facts in explaining a question is not immediately apparent when they are looked at in isolation, and only becomes more evident when they are considered as part of a larger explanation. Consider the question: ``\textit{George warms his hands by rubbing them. Which skin surface produces the most heat? (A) dry palms}''. The relevancy of ``\textit{1: as moisture of an object decreases, the friction of that object against another object increases}'' is clearer when 
``\textit{2: friction causes the temperature of an object to increase}'' is also known. Without the latter, someone (without world knowledge) might, for instance, regard 
facts about any other physical property that varies with moisture level as equally relevant to the question as the former, while they are not.

Second, by processing multiple facts at once, we can leverage BERT as a reading comprehension model, rather than as a retrieval model. The task requires not merely retrieving facts that seem relevant, like a search-engine would, but gathering a set of facts from which the answer to a simple science question can be inferred. 
That clearly requires more reasoning than a traditional retrieval task.
Research has shown that pre-trained transformers are able to infer knowledge from paragraphs of text, which is why they are more suited to handle this formulation of the task \cite{liu2019roberta,clark2019ftoa}.

Since we only need to be able to retrieve the facts with the highest probabilities, we can avoid computing normalized probabilities and instead compute scores (i.e., unnormalized probabilities). During training (next §) we do compute probabilities, in order to compute losses, but only over subsets of facts.

\subsubsection*{Training} \label{ssec:training}

\paragraph{Samples.}
The training input is encoded as $\bm{x} = [\bm{q} | \bm{f}^*_{1,...,N} | \bm{f}^{c} ]$, with $\bm{q}$ a query (question and answer), $\bm{f}^*_{1,...,N}$ a set of {gold} facts, and $\bm{f}^{c} \in \textit{vis}_k(\bm{q}_i, \bm{f}^*_{1,...,N})$ a candidate fact, which can be positive or negative, from the visible neighborhood of $[\bm{q} \mid \bm{f}^*_{1,...,N} ]$. We train our scoring function $f_\theta$ with stochastic gradient descent,
to output high scores for positive candidate facts and low scores for negative candidate facts.

Training samples for one $\bm{q}$ are constructed by first uniformly sampling a number $N\leq G$ of gold facts $\bm{f}^*_{1,...,N}$ from $\bm{q}$'s full set of gold facts. $N$ is itself uniformly sampled: $N \sim \mathcal{U}(0, G)$. The query and gold facts are concatenated into a prefix $\bm{p} = [\bm{q} | \bm{f}^*_{1,...,N}]$.
For one given prefix,
positive training samples $\bm{x}_p$ are constructed by concatenating $\bm{p}$ with all remaining visible gold facts.
Negative training samples $\bm{x}_n$ are constructed by concatenating the same $\bm{p}$ with a number of uniformly sampled visible negative facts\footnote{
Note that $\bm{f}^*_{1,...,N}$ in the prefix of both positive and negative training samples 
consists only of gold facts, and that all candidate facts $\bm{f}^{\cdot,c}$, gold or not, come from the visible neighborhood $\textit{vis}_k(\bm{p})$. 
}.
We either use all visible negatives, or sample a number until the minibatch is full.
The process is repeated: multiple prefixes are constructed for every query in $\mathcal{D}_{qa}$, and every prefix is appended with multiple visible, gold and negative facts. We construct multiple prefixes per $\bm{q}$ so that we have multiple training samples per $\bm{q}$, and so that the model is trained with different explanation lengths.

The prefix itself serves as a sample as well: the model is trained to score $\bm{p}$ highest when no more visible gold facts remain, i.e., when $\bm{f}^*_{1,...,N}$ contains all gold facts or when the remaining gold facts are not visible. During inference, $\bm{p}$ getting the highest score in an iteration is a stopping condition: the gathered set of facts is then considered complete.

\paragraph{Losses.}
Because classical maximum likelihood training for eq. \ref{eq:autoregr} would allow to backpropagate a loss only after all candidate facts have been considered, i.e., after 
$K_N$ forward passes, we resort to different loss functions. A simple loss that can be used is the pointwise binary cross-entropy loss (\textit{bXENT}), which considers each input example $\bm{x}$ individually and trains to correctly classify the candidate $\bm{f}^{c}$ as relevant or irrelevant. We propose to use the pairwise \textit{RankNet} loss \cite{burges2005ranknet}:

\begin{equation} \label{eq:ranknet}
	L(\bm{x}_p, \bm{x}_n; \theta) = - \log(\sigma(f_\theta(\bm{x}_{p}))-\sigma(f_\theta(\bm{x}_{n}))) \ ,
\end{equation}

Where $\sigma$ is the logistic sigmoid function, and $\bm{x}_{p}$ and $\bm{x}_{n}$ are samples in which $\bm{f}^{c}$ is a positive and negative fact, respectively. This loss is shown by %
\newcite{chen2009rankingmeasures} to maximize a lower bound on the MAP. To further amplify between-fact interactions in the gradient, we also use the conditional ranking variant of \textit{Noise-Contrastive Estimation} (NCE), which covers $>$2 facts at once \cite{ma2018nce,gutmann2010noise}:

\begin{equation} \label{eq:nce}
	L(\bm{x}_{1,...,B}; \theta) = - \log \frac{\exp({f}_\theta(\bm{x}_{1}) - \log P_n(\bm{f}^{c}))}{\sum_{j=1}^B \exp({f}_\theta(\bm{x}_{j}) - \log P_n(\bm{f}^{c}))} \ ,
\end{equation}

Where $\bm{x}_{1}$ is positive and $\bm{x}_{>1}$ are negative, $B$ is the batch size,
and $P_n$ is a negative sampling distribution over candidates: a uniform distribution over $\textit{vis}_k(\bm{p})$.
This loss has been used for training word embeddings as a more efficient approximation to the negative log-likelihood loss \cite{mnih2013learning,mikolov2013distributed}.
When training with NCE and RankNet, samples in one batch share a common prefix $\bm{p}$ and only differ in their candidate fact $\bm{f}^{c}$, so that all samples $\bm{x}$ in one loss term $L(\cdot,\theta)$ (eqs. \ref{eq:ranknet}-\ref{eq:nce}) only differ in $\bm{f}^{c}$ and hence the model is trained to score candidate facts and not prefixes.

The proposed training scheme -- training a model to predict the next gold element conditioned on previous gold elements -- is reminiscent of training text generation models with teacher forcing. A known weakness of teacher forcing is \textit{exposure bias} \cite{ranzato2016sequence}; models are conditioned on ground-truth data during training, as opposed to on their own outputs during inference. 
ARCF exhibits this discrepancy as well, which is why, during training, we try replacing a uniformly sampled amount of gold facts $\bm{f}^*_{1,...,N}$ (in the prefix) with uniformly sampled negative facts from $\textit{vis}_k(\bm{p})$. This feature is called `CN' later. Hence the model is trained to be more robust to mistakes it makes during inference. 
A similar technique was already proposed for text generation as \textit{scheduled sampling} \cite{bengio2015scheduled}.

\subsubsection*{Inference} \label{ssec:inference}

At inference, we incrementally build an explanation, i.e., a set of facts. The input follows the same encoding format: $\bm{x} = [\bm{q}|\bm{f}_{1,...,N}|\bm{f}^{c}]$ where $\bm{f}_{1,...,N}$ are previously selected facts (and not gold facts like in training).
At each iteration, we use the query $\bm{q}$ concatenated with already selected facts $\bm{f}_{1,...,N}$ as updated retrieval query, and rank all other {visible} facts $\bm{f}^c \in \textit{vis}_k(\bm{q} , \bm{f}_{1,...,N})$.
The highest scored fact is appended to the query for next iteration.
Note that the set of visible facts is extended with the neighborhood of the selected fact in each iteration.
The set of selected facts is considered to be complete when its cardinality $N$ equals $L$ or when the highest scored sample is the sample without candidate appended (see prev. §).
Multiple 
rankings are made; one with each intermediate set of facts and the question as retrieval query.
Algorithm \ref{alg:inference} shows the procedure in pseudocode.

\begin{wrapfigure}{r}{0.47\textwidth}
	\centering
	\begin{minipage}{0.47\textwidth}
		\begin{algorithm}[H]
		    \fontsize{9.5}{11}\selectfont
			\begin{algorithmic}
				\State \textbf{Input}: $\bm{q} \in \mathcal{D}_{qa}$, $\bm{f}_j \in \mathcal{D}_{f}$, neighborhoods $\textit{vis}_k$,
				\State \hspace{0.8cm} scoring function $f_\theta$, max length $L$
				\State $facts \gets  \emptyset$, $allscores \gets \emptyset$, 
				\State $candidates \gets  \textit{vis}_k(\bm{q})$, $l \gets 0$
				\While{$l < L$ \textbf{and} not stopping condition}
					\State $scores \gets \emptyset$
					\For{$j = 1 ...  |candidates| $} 
						\State $scores[j] \gets f_\theta( \, [ \bm{q} \, | \, facts \, | \,   candidates[j] \, ] \, )$
					\EndFor
					\State $facts \gets facts + [ \bm{f}_{\argmax(scores)} ] $
					\State $candidates \gets  \textit{vis}_k( \bm{q}; \, facts )$
					\State $allscores[l] \gets scores$,  $l \gets l + 1$
				\EndWhile
				\State \Return produce ranking($ facts, allscores $)
			\end{algorithmic}
			\caption{Inference procedure for one $\bm{q}$} \label{alg:inference}
		\end{algorithm}
	\end{minipage}
\end{wrapfigure}

When the stopping condition is met, we end up with an explanation $\hat{\bm{p}} = [\bm{q} | \bm{f}_{1,...,M}]$. To convert the result to a ranking, $ \bm{f}_{1,...,M}$ are ranked highest. Facts that were considered as candidate facts but not selected are ranked next; their relative order is determined by their scores in the last iteration. 
Next, all facts that were never considered are ranked by a simple metric like {tf-idf} distance from the computed explanation $\hat{\bm{p}}$.
We experimented with beam search procedures, where the beams were intermediate sets of facts. This did not improve performance on the validation set, so we do not consider it further.

This iterative 
fact selection 
bears similarity with how token-per-token text generation is usually performed with neural networks. Instead of computing a probability distribution over the vocabulary in one forward pass, 
our procedure requires a forward pass per score, i.e., per fact considered. To keep required resources for inference reasonable, only neighborhoods of facts are considered, instead of all facts, reducing the number of forward passes in one iteration from $D_f = 9707$ to $K_N$. Inference for a single $\bm{q}$ requires 
$L+\sum_{l=1}^{L} K_{l-1}=\mathcal{O}(L^2k)$ forward passes\footnote{
$L+\sum_{l=1}^{L} K_{l-1}\leq L+\sum_{l=1}^{L}l\cdot k=L+\frac{L}{2}(k+Lk)=\mathcal{O}(L^2k)$, with $L$ the max. number of iterations and $K$ defined in §\ref{ssec:fact-lands}: $K_l\leq (l+1)\cdot k$. 
}, with $L$ a chosen maximum number of iterations and $k$ the neighborhood size.
Parameter $k$ controls the trade-off between completeness and efficiency.

\section{Experiments} \label{sec:exp}

\subsection{Data \& Preprocessing} \label{ssec:data}

In the 2020 
version of the task $\mathcal{D}_f$ contains 9727 facts, and $\mathcal{D}_{qa}^{train}$, $\mathcal{D}_{qa}^{val}$, $\mathcal{D}_{qa}^{test}$ contain 2206, 496 and 1664 questions respectively \cite{xie2020worldtreev2}. The dataset has been extended w.r.t. the 2019 version of the task. For completeness, we also include results obtained with baselines and our models on the 2019 data. The 2019 data includes 902, 214 and 541 questions for training, validation, and testing respectively, along with 4950 facts \cite{jansen2018worldtreev1}.
We remove incorrect answers from $\bm{q}$ (like %
\newcite{das2019chains}), mark the correct answer with ``(answer)'' and the start of the gold facts with ``(explanation)''\footnote{ 
It is worth noting that technically, ARCF can perfectly run without the correct answer marked and without incorrect answers removed, although a performance drop on the explanation regeneration task can be expected.
}.
An example of a tokenized input sample could be ``[START] \textit{When does water start boiling? (answer) At 100\degree C. (explanation) This is a gold fact. This is another gold fact.} [SEP] \textit{This is a candidate fact that is gold or not} [SEP]''. Examples of $\bm{q} \in \mathcal{D}_{qa}$ and their explaining facts are shown in Appendix \ref{app:data-examples}. 

\subsection{Baselines}  \label{ssec:baselines}

As a simple baseline, we include a \textit{tf-idf} vector retrieval model, that ranks facts by cosine similarity between their and the $\bm{q}$'s {tf-idf} representation. We stem facts and $\bm{q}$, and remove stopwords before computing {tf-idf} vectors. 
Baseline \textit{single-fact} concatenates a $\bm{q}$ and a single candidate fact, and computes a relevance score for all $\bm{f}^c\in\mathcal{D}_f$ by encoding the concatenation $[\bm{q}\mid\bm{f}^c]$ with BERT and projecting the final layer's CLS-token embedding to a scalar with a linear layer \cite{das2019chains,nogueira2019passage}. The model is trained with binary cross-entropy (relevant or not).

The highest score in the 2019 competition was obtained by an ensemble of baselines \textit{single-fact} and \textit{path-rerank} \cite{das2019chains}. Model \textit{path-rerank} ranks facts for a $\bm{q}$ by first retrieving the top-$T$ facts with the \textit{tf-idf} retriever from above.
This initial top-$T$ set is extended with all facts that have $\geq$1 words in common with one of the top-$T$ facts. Next, all combinations of up to $C = 2$ facts are taken from this extended set. A relevance score is computed for all combinations (\textit{chains}),
in the same way as in \textit{single-fact} or our models: concatenate $\bm{q}$ and the $C$ facts, encode the concatenation with BERT and project the final CLS-token embedding to a score $s$ with a linear layer.
A fact's relevance score is the maximum score of any chain it appears in. 
The binary cross-entropy loss is used for training the model.
We use the implementation of \newcite{das2019chains} for the \textit{single-fact} and \textit{path-rerank} baselines\footnote{Code available at \url{https://github.com/ameyagodbole/multihop_inference_explanation_regeneration}.
}.

\paragraph{Complexities.} %
\newcite{das2019chains} used \textit{single-fact} and \textit{path-rerank} for the smaller 2019 dataset, with fewer facts and fewer $\bm{q}$. They already noted that \textit{single-fact} is not scalable to a large corpus of facts: for the 2020 data, $D_f \approx 10$K forward passes are required to solve a single $\bm{q}$ during inference. One epoch trained with bXENT consists of 21M samples ($D_{qa}^{train}\cdot D_f$, trained 3 epochs).
The \textit{path-rerank} model uses $T=25$ for training, which generates 7k chains of facts per $\bm{q}$, resulting in 16M training samples (trained 1 epoch). Using $T=50$ during inference results in 16k chains (hence forward passes) per $\bm{q}$. This number can be controlled by setting $T$, but setting $T$ too small would leave too many relevant facts out of consideration. In contrast, the neighborhood in our method is less restrictive as it depends on selected facts and thus expands progressively as more facts are selected.

\subsection{Experiments} \label{ssec:experiments}

We implemented our algorithm and baselines using PyTorch and the Transformers library\footnote{
Our code and trained models are publicly available at \url{https://github.com/rubencart/LIIR-TextGraphs-14}.
} \cite{paszke2019pytorch,wolf2019huggingface}.
The \textit{tf-idf} baseline was implemented with SciKit Learn \cite{pedregosa2011scikit}.
To keep comparisons fair,
all results on 2019 data (baselines and ARCF) are obtained by finetuning the publicly available pre-trained \textit{bert-base-uncased} (since this model is used in %
\newcite{das2019chains}). To reduce resource usage, all models on 2020 data were finetuned from the smaller pre-trained \textit{distilroberta-base}\footnote{
When we finetuned \textit{bert-base-uncased} on the 2020 data or \textit{distilroberta-base} on the 2019 data we obtained similar 
results.
}.
It can reasonably be expected that using bigger or more advanced pre-trained models would further improve results.
We ran experiments on 1 16GB Nvidia Tesla P100 GPU. We used the Adam optimizer \cite{kingma2015adam}, with learning rate \num{2e-5} and linear LR decay. We append samples to minibatches until they reach 5000 tokens. An overview of used hyperparameters can be found in appendix \ref{app:hyperparams}. 
For training we set 
neighborhood size $k=180$ ($L$ only impacts inference), for inference we set the maximum and minimum number of iterations $L=9$, $L_{min} = 3$, and $k=290$.
Some hyperparameters were taken from %
\newcite{das2019chains}, while others were tuned manually.

In the remainder of this section, \textbf{ARCF} denotes our proposed method, \textbf{SF} refers to the \textit{single-fact} baseline, \textbf{PR} is the \textit{path-rerank} baseline, \textbf{CN} means `conditioned on negatives', \textbf{S2} means `rank scored but not selected facts 2\textsuperscript{nd}', \textbf{R3} means `rank rest 3\textsuperscript{rd}' (as opposed to omitting them altogether, see §\ref{ssec:inference}).

\subsection{Results and Discussion} \label{ssec:results}

\paragraph{Test set results.}
Tables \ref{tab:test-maps},\subref{tab:test-times} show results on the hidden 2019 and 2020 test sets\footnote{
Leaderboard of 2020 competition: 
\url{https://competitions.codalab.org/competitions/23615} (26/10/2020), 2019 competition: \url{https://competitions.codalab.org/competitions/20150} (26/10/2020).
}, total training time and inference time per sample, for the baselines and ARCF. 
The test scores are obtained by models that got the highest validation score of 5 training runs with different random seeds, while the training times are averaged over these 5 runs. As can be seen, ARCF outperforms the baselines both in terms of obtained MAP and efficiency\footnote{
\newcite{xie2020worldtreev2} report a MAP score of 0.52 instead of 0.4992 on the 2020 test set with a \textit{single-fact} BERT baseline,
but since the publication of their paper the 2020 dataset (incl. the test set) has been updated. They also use BERT instead of RoBERTa, and possibly different hyperparameters.
}. The highest 2020 MAP is 0.5815, which put us at the second place in the online competition.
Appendix \ref{app:output-samples} shows examples of validation set questions and predicted facts. 

On the 2020 test set, all our models obtain a higher MAP than all baselines. This is not the case on the 2019 data, %
which suggests that ARCF benefits more from additional training data. 
Including $>$2 facts in one loss term with NCE shows no benefit compared to the pairwise RankNet loss. 
Conditioning on negatives has no significant impact. 
Five 2020 test set evaluations of ARCF and SF show that the difference in scores is statistically significant: ARCF with RankNet scores \textit{higher} on average than SF (1-tailed indep. t-test, $p<0.001$).

\begin{table}
\centering
\fontsize{9}{11}\selectfont
\begin{subtable}[t]{0.5\textwidth}
\begin{tabular}[t]{p{0.34\textwidth}p{0.15\textwidth}>{\centering}p{0.18\textwidth}>{\centering\arraybackslash}p{0.18\textwidth}}
 Algorithm &  Loss          & MAP  2020  & MAP 2019 \\ \toprule %
 TF-IDF         & 		    & 0.3743    &  0.3870     \\
 SF    & bXENT	    & 0.4992    &  0.5574     \\ %
 PR& bXENT 	& 0.4629    &  0.5313     \\ 
 ENSEMBLE       & bXENT     & 0.5081    &  0.5625    \\ \midrule
 
 ARCF   & RankNet   & \textbf{0.5815}  &  \textbf{0.5707}     \\ %
 \quad with CN                 & RankNet   & 0.5810    &  0.5575 \\
 ARCF	    & NCE 	    &  0.5728    &   0.5634        \\
 \quad with CN                 & NCE       &  0.5759    &       0.5691  \\ %
\end{tabular}
\caption{} \label{tab:test-maps}
\vspace{0.3cm}
\begin{tabular}[t]{p{0.34\textwidth}p{0.15\textwidth}>{\centering}p{0.18\textwidth}>{\centering\arraybackslash}p{0.18\textwidth}}
Algorithm &  Loss          & test 2020  & val 2020 \\ \toprule
ARCF                & RankNet     & 0.5815 & 0.5931 \\
\quad w bXENT                           & bXENT     &  $-$0.0082  &   $-$0.0060  \\    %
\quad w SF train    & bXENT     & $-$0.102 & $-$0.101  \\
\quad w SF inf      & RankNet   & $-$0.037  & $-$0.041 \\
\quad w/o prefix, neighb.   & RankNet   & $-$0.053 & $-$0.054 \\ \midrule
\quad w/o R3 	                & RankNet   &  $-$0.0004  & $-$0.0001 \\ %
\quad w/o R3, S2 	            & RankNet   &  $-$0.072  & $-$0.079 \\ %
	\end{tabular}

	\caption{} \label{tab:abl-maps}
\end{subtable}
\begin{subtable}[t]{0.49\textwidth}
\centering
    \begin{tabular}[t]{lcc}
    Algorithm       & train T (H) &  inf T (s$/ \bm{x}$) \\ \toprule %
    {SF}   & 56.3         & 18.4 \\
    {PR}    & 16.2     	& 31.8 \\
    ARCF %
    & 5.7     &  9.6 \\
    \end{tabular}
    \caption{} \label{tab:test-times}
\vspace{0.5cm}
    \begin{tabular}{l@{\hspace{1\tabcolsep}}c@{\hspace{1\tabcolsep}}c@{\hspace{1\tabcolsep}}c@{\hspace{1\tabcolsep}}c@{\hspace{1\tabcolsep}}c}
    $k$       & 90  & 130 & 180 & 210 & 290 \\ \toprule
    MAP         &.567 & .577 & .581 & .581 & .581 \\    %
    T (s$/\bm{x}$) &2.5&3.6&4.9&5.7&7.8\\%& 19:58 & 28:58 & 40:16 & 46:56 & 1:03:43 \\
    \end{tabular}
    \caption{} \label{tab:k}
\vspace{0.5cm}
    \begin{tabular}{l@{\hspace{1\tabcolsep}}c@{\hspace{1\tabcolsep}}c@{\hspace{1\tabcolsep}}c@{\hspace{1\tabcolsep}}c@{\hspace{1\tabcolsep}}c@{\hspace{1\tabcolsep}}c@{\hspace{1\tabcolsep}}c}
    $L$       & 2 & 4 & 6 & 8 & 10 & 12 & 14 \\ \toprule
    MAP         & .561 & .576 & .581 & .581 & .581 & .581 & .582 \\%& 0.5776 & 0.5840 & 0.5870 & 0.5885 & 0.5897 & \\
    T (s$/\bm{x}$) & 0.43 & 1.4 & 2.9 & 4.9 & 7.6 & 11.0 & 15.1 \\%& 11:38 & 23:32 & 40:16 & 1:02:20 & 1:30:28 \\
    \end{tabular}
    \caption{} \label{tab:L}
\end{subtable}

\caption{(\subref{tab:test-maps}) Mean average precision (MAP) of baselines (upper) and ARCF (bottom) on the 2020 and 2019 test sets. 
(\subref{tab:abl-maps}) Ablations on the 2020 test and validation sets.
(\subref{tab:test-times}) Training time in hours and inference time in seconds/sample.
(\subref{tab:k})-(\subref{tab:L}) Impact of $k$ (with $L=8$) and $L$ (with $k=180$) on 2020 test MAP and inference time.
} \label{tab:test-results}
\end{table}

\paragraph*{Ablation study.} 
As ARCF consists of several components, we perform an ablation study on the 2020 test and validation sets. Results are shown in table \ref{tab:abl-maps}. First, ARCF was trained with the pointwise \textbf{bXENT} loss (but still with prefixes and neighborhoods). The MAP drops,
showing that pairwise information in the gradients improves learning w.r.t. pointwise information.
Second, ARCF is compared to \textbf{ARCF w SF train}, which is trained like SF and uses algorithm \ref{alg:inference} for inference. The large drop in MAP indicates that algorithm \ref{alg:inference} only works for inference if the model is trained accordingly. 
Next, \textbf{ARCF w SF inf} is trained like ARCF and evaluated like SF. This result shows that ARCF training still improves performance w.r.t. SF even when facts are scored individually, and it shows that algorithm \ref{alg:inference} improves performance.
\textbf{ARCF w/o prefix, neighb} is trained with the pairwise RankNet loss but without prefixes and neighborhoods. Minibatches each contain 1 positive and $B-1$ negatives that are uniformly sampled from all $D_f - G$ negative facts. Evaluation is carried out like for SF. The score drops almost 10\%, which emphasizes both the need for informative negatives when training with a pairwise loss, which our neighborhoods provide by returning nearby facts, and the importance of training with prefixes. It also affirms the gain of algorithm \ref{alg:inference} for inference.
Leaving facts that were never scored out of the ranking (\textbf{w/o R3}) has negligible impact. Ignoring scored but unselected facts too (\textbf{w/o R3, S2}) cuts performance by $>$10\%: including them is thus essential. Visible facts are scored anyway, so including them all in the ranking comes at no additional cost, except for once sorting them based on their score.
We also trained a randomly initialized version of \textit{distilroberta-base} (instead of pre-trained). Although we did not extensively tune hyperparameters, the maximum obtained test MAP was about 25\% of pre-trained models.

\paragraph{Impact of neighborhood size $k$ and maximum explanation length $L$.} As tables \ref{tab:k}-\subref{tab:L} show, MAP increases with $k$ and $L$, before flattening around $k=180,L=8$. Inference time per sample increases approximately quadratically with $L$ and linearly with $k$, which is in line with the number of FW passes for ARCF inference growing with $\mathcal{O}(L^2k)$. Fig. \ref{fig:L-k-plot} in appendix \ref{app:extra-plots} also shows this. Table \ref{tab:k} shows that we could have taken $k=180$ for inference, with virtually no loss in MAP and almost 40\% speedup, but the $k=290$ we used showed better validation results (likewise for $L=6$).

\paragraph{Performance on subsets of facts.}

\begin{figure}
    \centering
    \begin{subfigure}{0.3\textwidth}
        \centering
        \includegraphics[width=1.0\textwidth]{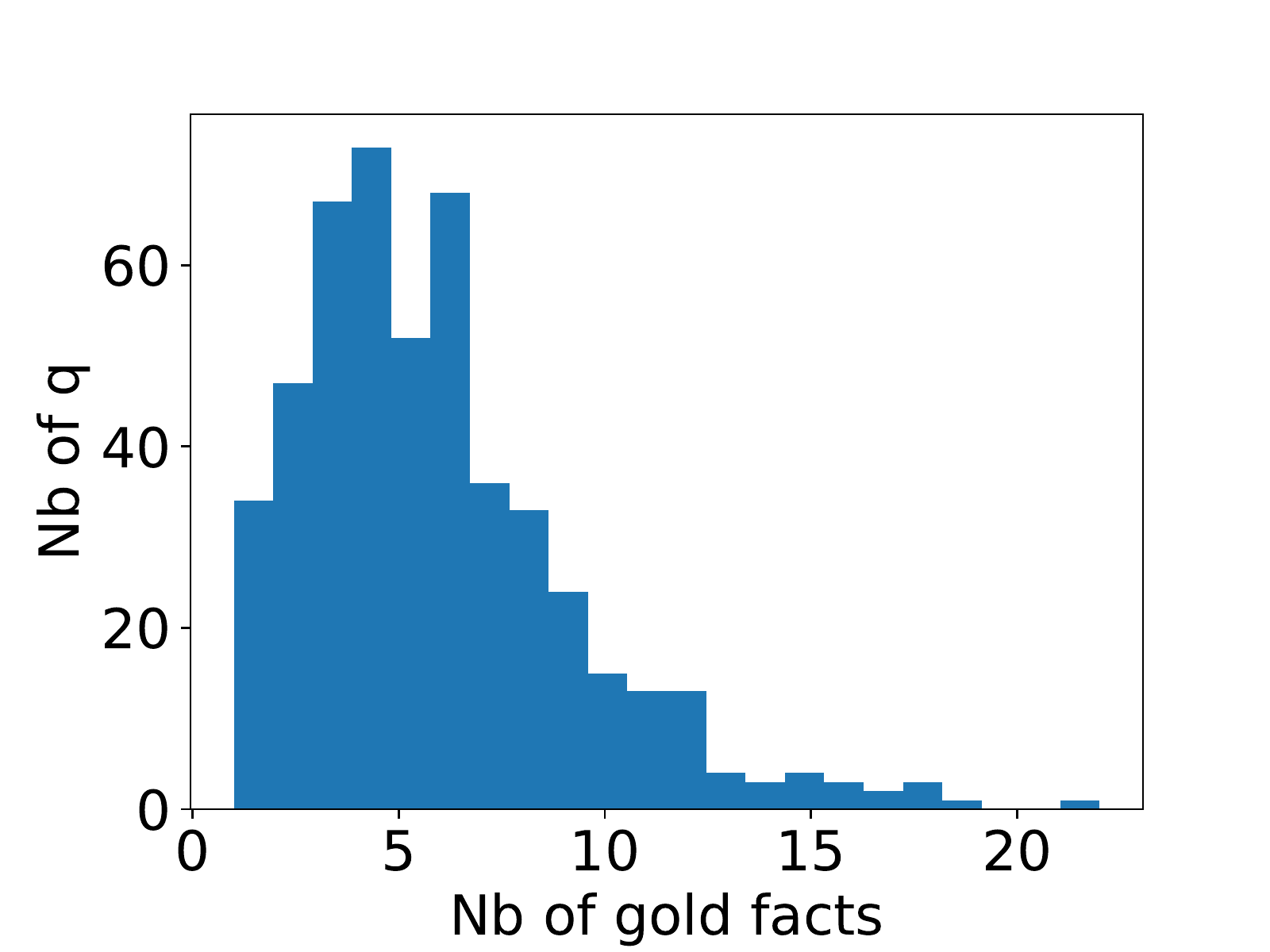}
        \caption{} 
        \label{fig:bd-lens}
    \end{subfigure}
    \begin{subfigure}{0.3\textwidth}
        \centering
        \includegraphics[width=1.0\textwidth]{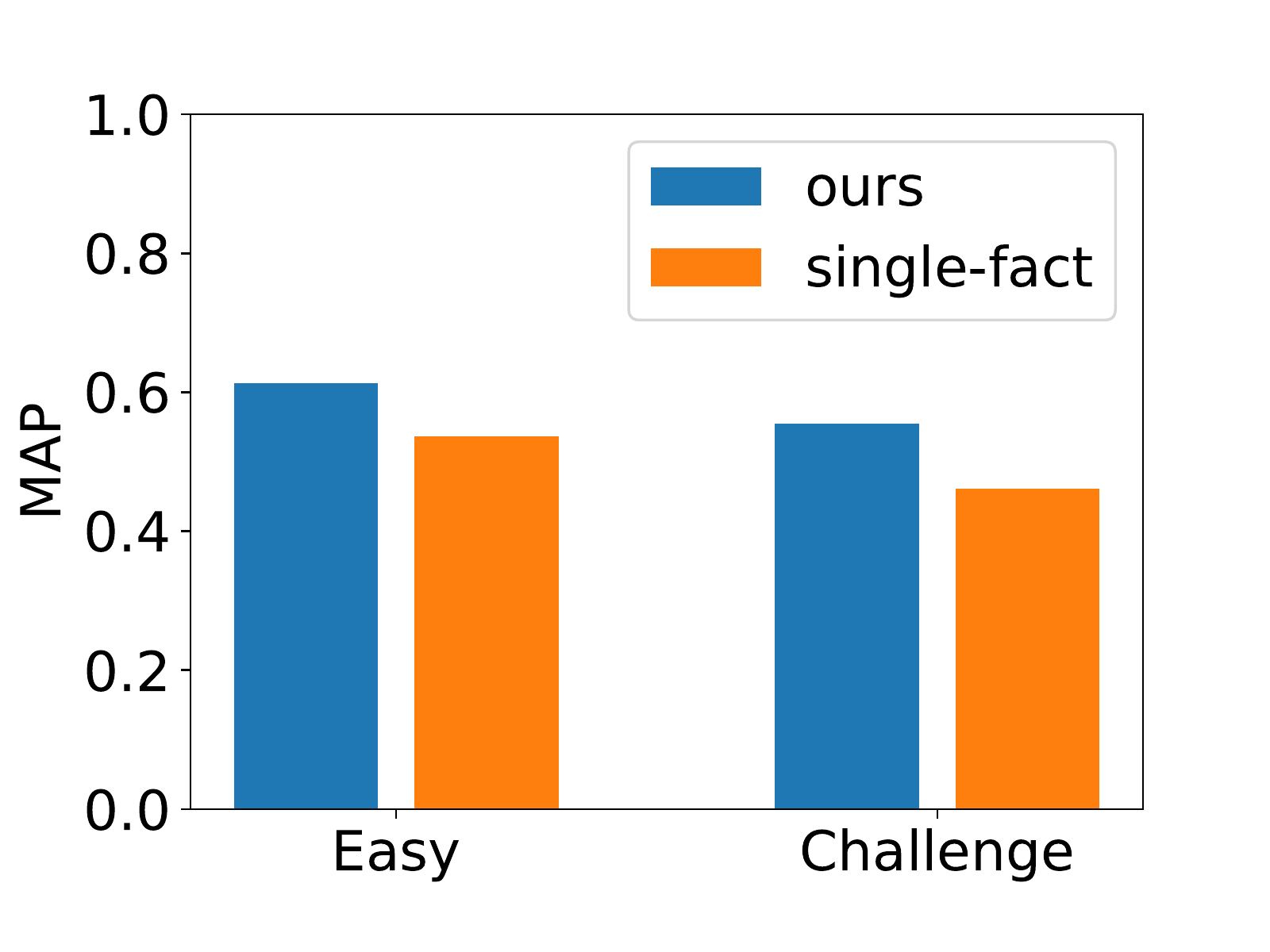}
        \caption{} 
        \label{fig:bd-lvl}
    \end{subfigure}
    \begin{subfigure}{0.3\textwidth}
        \centering
        \includegraphics[width=1.0\textwidth]{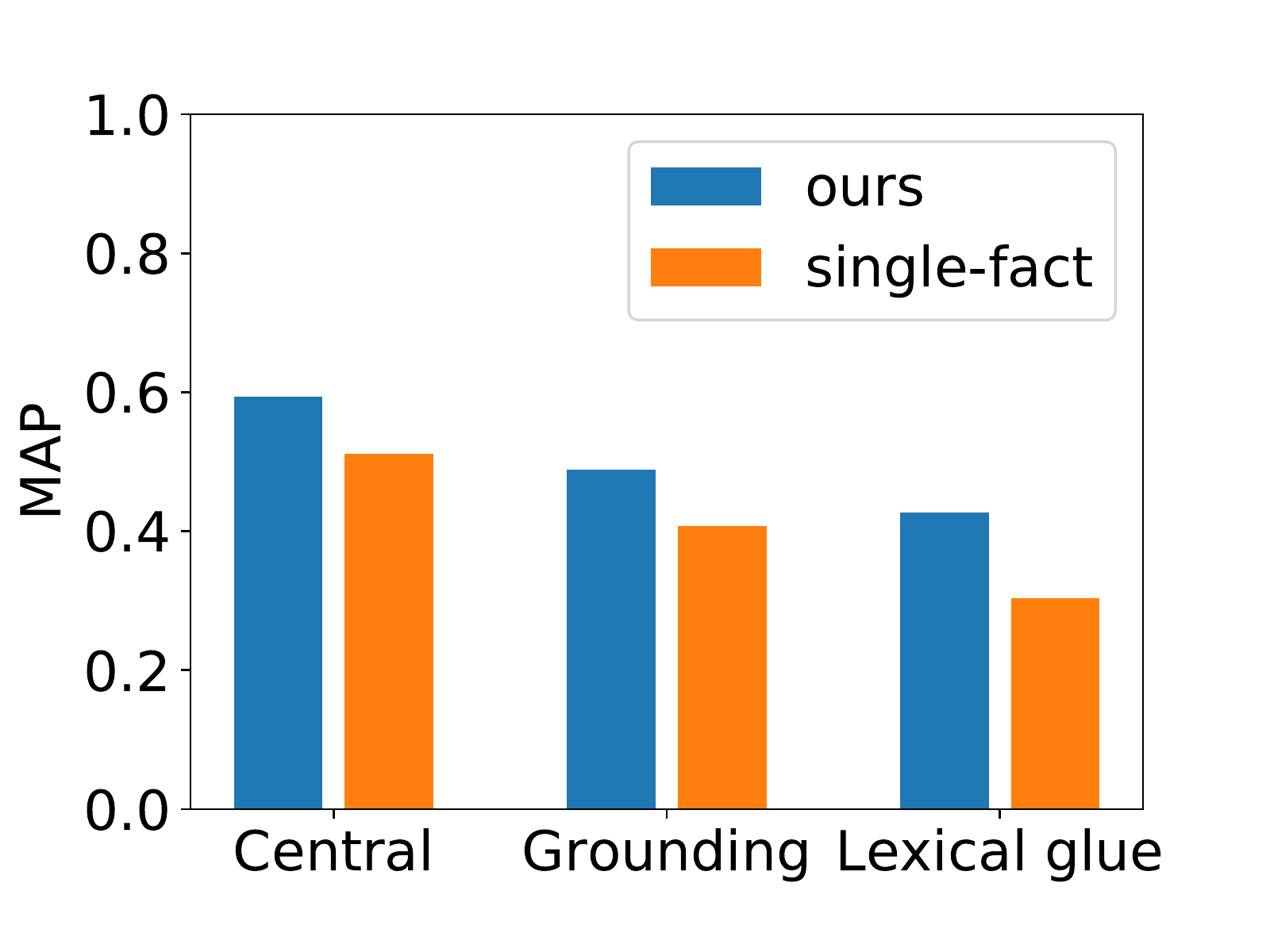}
        \caption{} 
        \label{fig:bd-role}
    \end{subfigure}
    \begin{subfigure}{0.3\textwidth}
        \centering
        \includegraphics[width=1.0\textwidth]{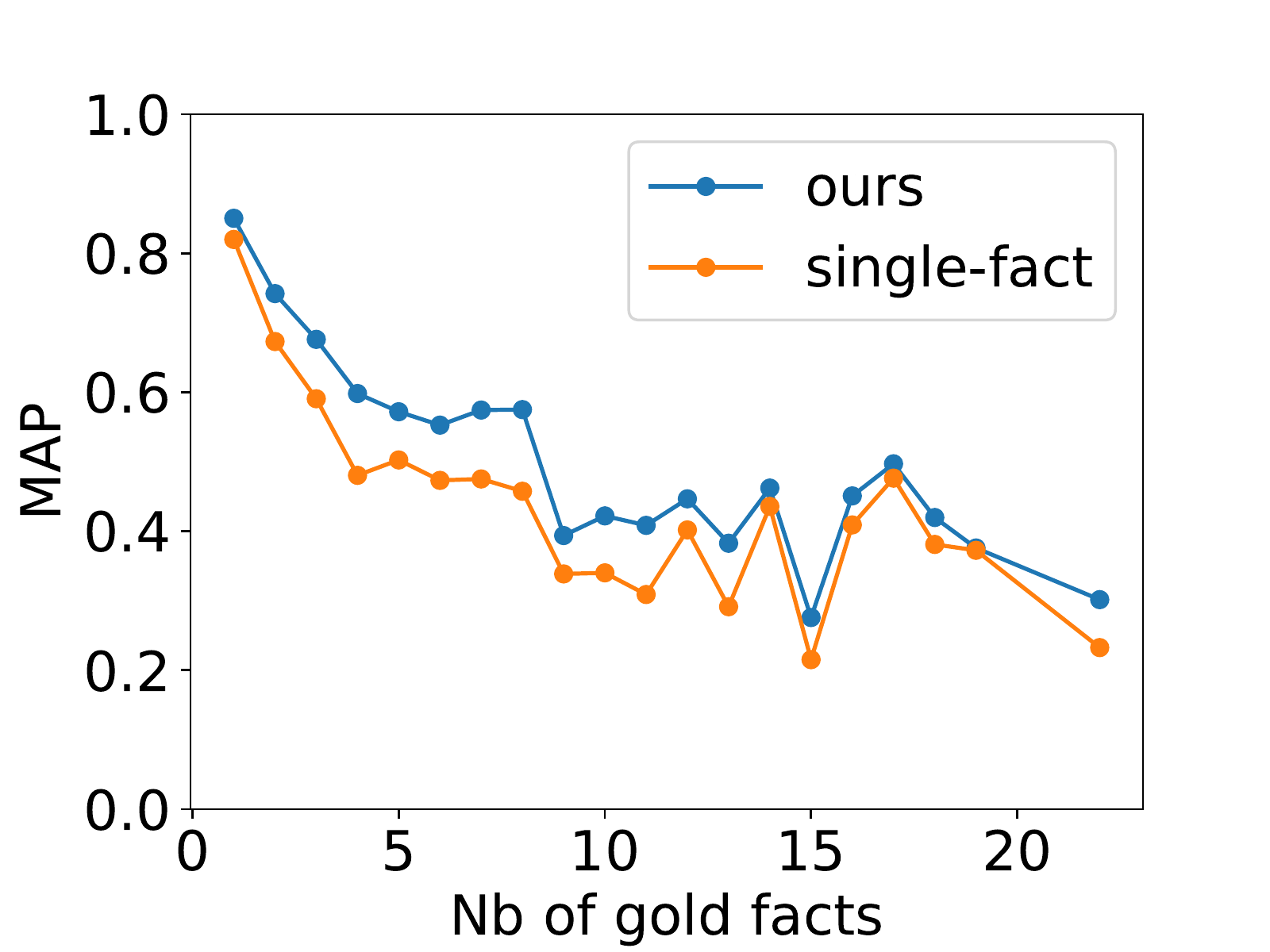}
        \caption{} 
        \label{fig:bd-maplen}
    \end{subfigure}
    \begin{subfigure}{0.3\textwidth}
        \centering
        \includegraphics[width=1.0\textwidth]{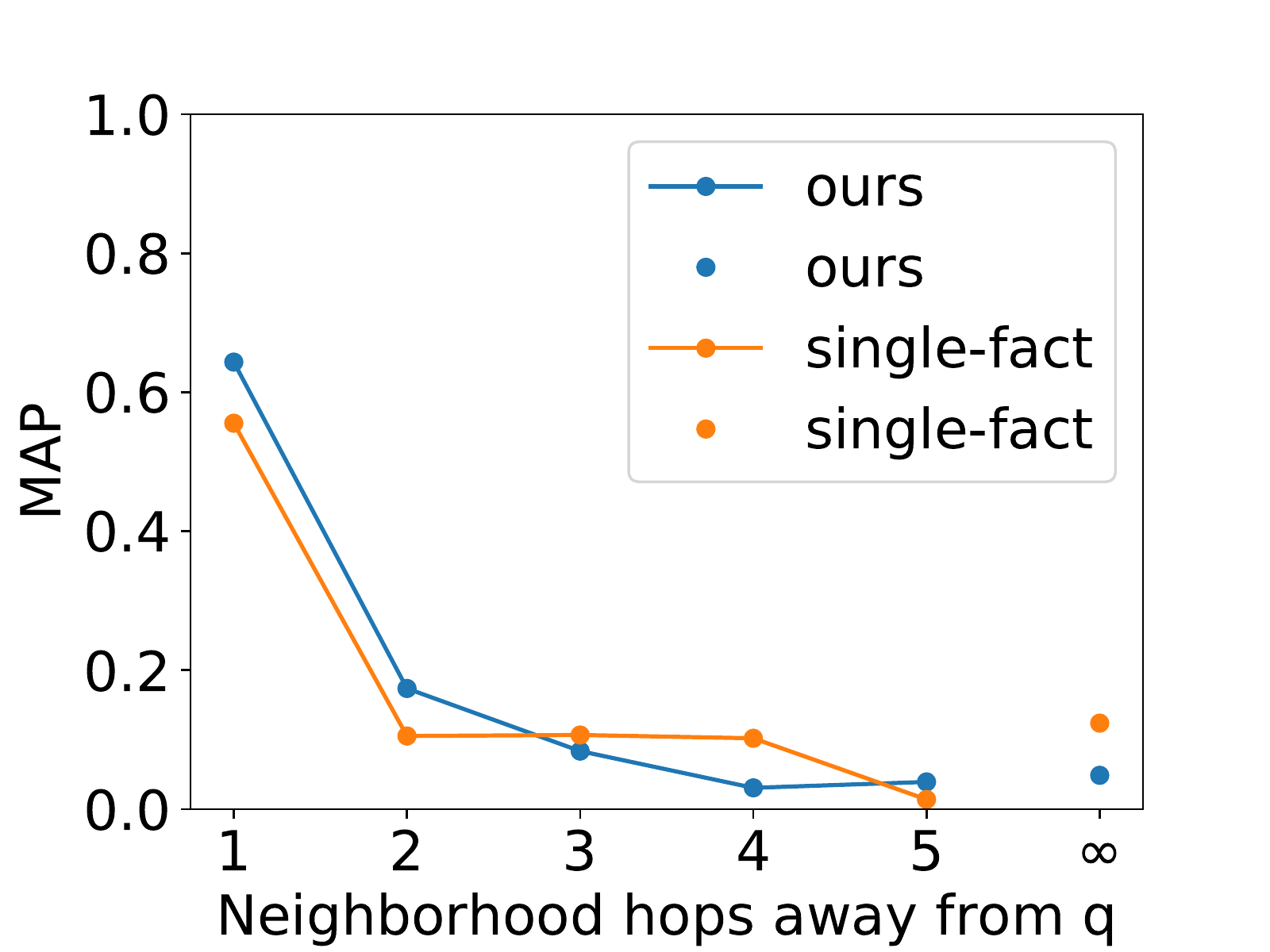}
        \caption{} 
        \label{fig:bd-kh}
    \end{subfigure}
    \begin{subfigure}{0.3\textwidth}
        \centering
        \includegraphics[width=1.0\textwidth]{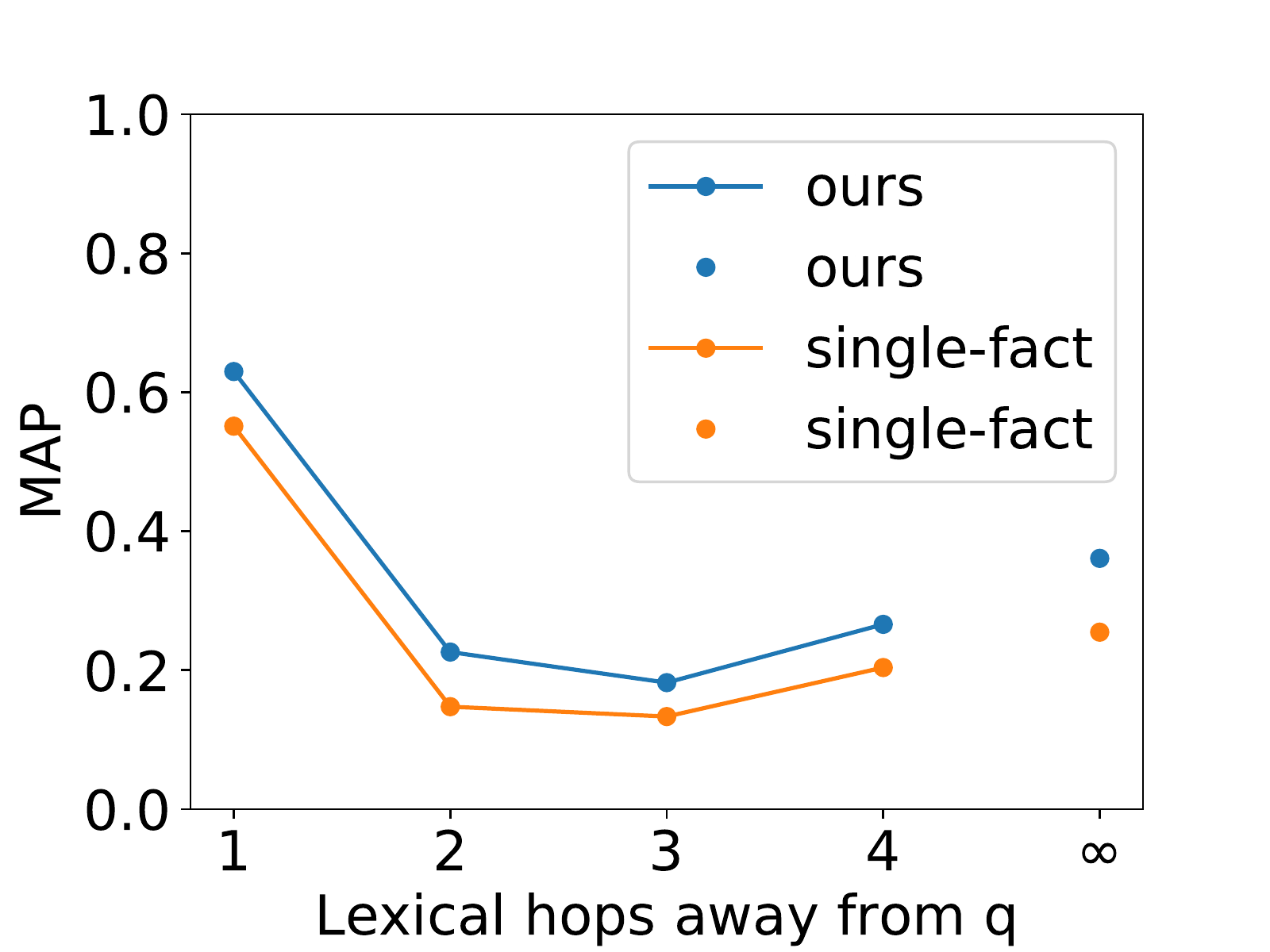}
        \caption{} 
        \label{fig:bd-lh}
    \end{subfigure}
    \caption{MAP of ARCF vs. \textit{single-fact} on subsets of facts.} 
    \label{fig:breakdown}
\end{figure}

Since the leaderboards only return a scalar test MAP,
we run a number of experiments on the smaller validation set. 
The results 
are therefore only indicative. %
Fig. \ref{fig:bd-lens} shows the number of $\bm{q} \in \mathcal{D}_{qa}^{val}$ for different numbers of corresponding gold facts ($G$). 
Fig. \ref{fig:bd-lvl}-\subref{fig:bd-role} show MAP scores of ARCF (RankNet)
vs. the SF baseline on subsets of facts as marked in the dataset. 
ARCF increases MAP on the Challenge subset with an absolute \%10\footnote{
All questions in the dataset are marked by annotators with either `Easy' or `Challenge' (all $\bm{q}$ are one of the two).
}.
It
significantly improves retrieval of all roles, but most of lexical glue and grounding facts.
These are facts that support central facts, which might be easier to detect when using the context of other facts as our model does. %
This is a useful improvement, as reasoning of explanations that contain lexical glue and grounding facts is easier to understand\footnote{Examples in appendices \ref{app:data-examples} and \ref{app:output-samples}.}.

Figs. \ref{fig:bd-kh}-\subref{fig:bd-lh} show the MAP for facts that need an increasing number of hops to reach (from $\bm{q}$ or a fact to another fact, only via gold facts).
In fig. \ref{fig:bd-kh} the hops are always to a fact in the current neighborhood $\textit{vis}_k(\cdot)$, in fig. \ref{fig:bd-lh} to facts with lexical overlap as seen in \newcite{das2019chains}.
The `$\infty$' on the x-axis shows the precision for facts that cannot be reached from $\bm{q}$ in any number of hops (they can still be correctly retrieved by our method, but only via a negative fact). 
Surprisingly, the MAP of ARCF drops below that of SF for facts that are $3-4$ neighborhood hops away. 
When lexical overlap hops are considered, ARCF performs better than the baseline for `farther' facts.
The precision values for figure \ref{fig:bd-role}-\subref{fig:bd-lh} are computed as by %
\newcite{jansen2019textgraphs}, by first removing gold facts that have another role (or are not exactly $h$ hops away)
both from the gold set and from the predicted ranking, and then computing the MAP of the remaining predicted ranking w.r.t. the remaining gold set\footnote{
E.g., to compute the MAP for central facts, gold grounding and lexical glue facts are removed from the predicted ranking, so they do not influence the central fact MAP. The MAP of the remaining predictions w.r.t. the gold central facts is then computed.
}.

Fig. \ref{fig:bd-maplen} shows the MAP for $\bm{q}$'s with different numbers of gold facts $G$ (the point at $x=2$ shows the MAP on those $\bm{q}$ that have a total of $G=2$ gold facts). ARCF gives a consistent improvement over the baseline for all $G$.

\section{Conclusion} \label{sec:conclusion}

\paragraph{Future work.} Future work might expand on our approach by finding alternative methods to evaluate a fact w.r.t. other facts in an iterative inference procedure, or by designing better fact-question neighborhood methods. %
Additionally, the performance of our method when the correct answer to a question is not given could be evaluated.
Future work could then infer the answer from retrieved facts in a downstream QA setting. 
Finally, one could improve the method by considering to \textit{remove} earlier chosen facts from the intermediate set of selected facts.

\paragraph{} We have proposed a new method to retrieve relevant facts for an explanation regeneration task by iteratively evaluating candidate facts with respect to previously selected facts using a learning-to-rank approach. We have successfully evaluated our method on the Textgraphs 2019 and 2020 datasets and have performed several ablation experiments. We have analyzed time complexity of our method and the performance on different subsets of facts.
By selecting the nearest facts by similarity between {tf-idf} vectors, considering not just the question but also 
already selected facts, only a subset of facts are considered at each step, and ARCF outperforms previous state-of-the-art methods at a higher efficiency. 

\section*{Acknowledgements}
The research leading to this paper received funding from the Research
Foundation Flanders (FWO) under Grant Agreement No. G078618N and from
the European Research Council (ERC) under Grant Agreement No. 788506. 
The Flemish Supercomputer Center (VSC) provided hardware and GPUs.

\clearpage
\bibliographystyle{coling}
\bibliography{references.bib}

\clearpage
\begin{appendices}
    
\section{Examples of TextGraphs-2020 data} \label{app:data-examples}

\begin{table}[H]
\centering
\fontsize{9.5}{11}\selectfont
\begin{tabular}[t]{l p{0.7\textwidth}} 
\toprule
Question 1 & The influence of the Moon on the tides on Earth is greater than that of the Sun. Which best explains this? (answer) The Moon is closer to Earth than the Sun. \\ \midrule
Fact 0 - Role central & the gravitational pull of the Moon  on Earth's oceans causes the tides \\
Fact 1 - Role central & as distance from an object decreases , the pull of gravity on that object increases \\
Fact 2 - Role grounding & closer means lower; less; a decrease in distance \\
Fact 3 - Role grounding & a moon is a kind of celestial object; body \\
Fact 4 - Role central & the moon is the celestial object that is closest to the  Earth \\
Fact 5 - Role grounding & the Sun is a kind of star \\
Fact 6 - Role grounding & a star is a kind of celestial object; celestial body \\
Fact 7 - Role central & the Moon is the celestial object that is  closer to the  Earth than the Sun \\
Fact 8 - Role grounding & Earth is a kind of planet \\
Fact 9 - Role grounding & a planet is a kind of celestial object; body \\
Fact 10 - Role lexglue & cause is similar to influence \\
Fact 11 - Role lexglue & gravity means gravitational pull; gravitational energy; gravitational force; gravitational attraction \\ \bottomrule
\end{tabular}

\vspace{1cm}

\begin{tabular}[t]{l p{0.7\textwidth}} 
\toprule
Question 2 & A student placed an ice cube on a plate in the sun. Ten minutes later, only water was on the plate. Which process caused the ice cube to change to water? (answer) melting.  \\ \midrule
Fact 0 - Role central & melting means matter; a substance changes from a solid into a liquid by increasing heat energy  \\
Fact 1 - Role grounding & an ice cube is a kind of solid  \\
Fact 2 - Role grounding & water is a kind of liquid at room temperature  \\
Fact 3 - Role central & water is in the solid state , called ice , for temperatures between 0; -459; -273 and 273; 32; 0 K; F; C  \\
Fact 4 - Role lexglue & heat means heat energy  \\
Fact 5 - Role lexglue & adding heat means increasing temperature  \\
Fact 6 - Role central & if an object; a substance; a location absorbs solar energy then that object; that substance will increase in temperature  \\
Fact 7 - Role central & if an object; something is in the sunlight then that object; that something will absorb solar energy  \\
Fact 8 - Role central & the sun is a source of light; light energy called sunlight  \\
Fact 9 - Role lexglue & to be in the sun means to be in the sunlight  \\
Fact 10 - Role central & melting is a kind of process  \\
 \bottomrule
\end{tabular}
\caption{Example $\bm{q} \in \mathcal{D}_{qa}^{val}$ with explaining gold facts and their roles. The wrong multiple choice answer options have already been removed.
} \label{tab:datset-samples}
\end{table}

\section{Hyperparameters} \label{app:hyperparams}

All ARCF models were trained with an L2 weight decay coefficient of $0.01$. We tried training baselines \textit{single-fact} and \textit{path-rerank} with the same weight decay, but validation and test set results were lower than without weight decay.

Tables \ref{tab:hp-train} and \ref{tab:hp-inf} show the hyperparameters we used for ARCF training and inference. Hyperparameters for training with different loss functions are largely the same, only the number of epochs trained might differ with 1 or 2.
When conditioning on negatives (CN) was used for training, we first trained for 2 epochs as normal and then started replacing a uniformly sampled proportion between 0.0 and 0.3 of gold facts in the prefix $\bm{f}_{1,...,N}^*$ by uniformly sampled negatives from the visible neighborhood of the prefix as it was before the replacement $\textit{vis}_k(\bm{p})$.
Parameter $L$ is only relevant for inference, as it is only used by algorithm \ref{alg:inference}.

\begin{table}[h]
\centering
\fontsize{10}{11}\selectfont
\begin{subtable}[t]{0.50\textwidth}
\centering
	\begin{tabular}[t]{lc} 
	Hyperparam & value \\ \toprule
    LR & \num{2e-5} \\
    Epochs & 4 \\
    L2 weight decay & 0.01 \\
    ADAM $\epsilon$ & \num{1e-8} \\
    ADAM $\beta_1$ & 0.9 \\
    ADAM $\beta_2$ & 0.999 \\
    $k$ & 180 \\
    Tokens in minibatch & 5000 \\
    
	\end{tabular}

	\caption{} \label{tab:hp-train}
\end{subtable}
\begin{subtable}[t]{0.49\textwidth}
\centering
    \begin{tabular}[t]{lc}
    Hyperparam & value \\ \toprule
    $k$ & 290 \\
    $L$ & 9 \\
    $L_{min}$ & 3 \\
    Tokens in minibatch & 24k \\
    R3 & tf-idf \\
    S2 & true \\
    \end{tabular}
    \caption{} \label{tab:hp-inf}
\end{subtable}

\caption{Hyperparameters used for ARCF training (with RankNet) and inference.
} \label{tab:hyperparams}
\end{table}

\section{Examples of validation set predictions} \label{app:output-samples}

Table \ref{tab:pred-samples} shows the 15 highest ranked facts by ARCF (RankNet) for the two validation questions in table \ref{tab:datset-samples}. The first column of each row for gold facts (which are correctly ranked high) is \textcolor{teal}{colored blue}. As can be seen in the predictions for Question 1, some wrongly predicted facts are clearly related to the question but not necessary for explaining the answer (e.g. Fact 0). While others (like Fact 2) could actually be used for explaining the question as well. Fact 2 for question 1 in table \ref{tab:pred-samples} could reasonably take the place of gold Fact 7 for the same question in table \ref{tab:datset-samples}.

\begin{table}[h]
\centering
\fontsize{9.5}{11}\selectfont
\begin{tabular}[t]{l p{0.8\textwidth}} 
\toprule
Question 1 & The influence of the Moon on the tides on Earth is greater than that of the Sun. Which best explains this? (answer) The Moon is closer to Earth than the Sun. \\ \midrule
\textcolor{teal}{Fact 0} & cause is similar to influence. \\ 
Fact 1 & as the gravitational pull of the moon on the Earth decreases , the size of the tides on Earth decrease. \\ 
Fact 2 & the Moon is closer to the Earth than the Sun. \\ 
\textcolor{teal}{Fact 3} & closer means lower; less; a decrease in distance. \\ 
Fact 4 & as the gravitational pull of the moon on the Earth decreases , the size of the tides on Earth decrease. \\ 
\textcolor{teal}{Fact 5} & gravity means gravitational pull; gravitational energy; gravitational force; gravitational attraction. \\ 
\textcolor{teal}{Fact 6} & as distance from an object decreases , the pull of gravity on that object increases. \\ 
Fact 7 & as distance from an object increases , the pull of gravity on that object decrease. \\ 
Fact 8 & an increase is the opposite of a decrease. \\ 
Fact 9 & as the distance from an object increases , the force of gravity on that object will decrease. \\ 
\textcolor{teal}{Fact 10} & a moon is a kind of celestial object; body. \\ 
\textcolor{teal}{Fact 11} & the gravitational pull of the Moon  on Earth's oceans causes the tides. \\ 
Fact 12 & the gravitational pull of the Sun on Earth's oceans causes the tides. \\ 
Fact 13 & less is similar to decrease. \\ 
Fact 14 & to lower means to decrease.  \\ 
\bottomrule
\end{tabular}

\vspace{1cm}

\begin{tabular}[t]{l p{0.8\textwidth}} 
\toprule
Question 2 & A student placed an ice cube on a plate in the sun. Ten minutes later, only water was on the plate. Which process caused the ice cube to change to water? (answer) melting.  \\ \midrule
\textcolor{teal}{Fact 0} & melting is a kind of process.  \\
\textcolor{teal}{Fact 1} & melting means matter; a substance changes from a solid into a liquid by increasing heat energy.  \\
\textcolor{teal}{Fact 2} & an ice cube is a kind of solid.  \\
Fact 3 & water is a kind of substance.  \\
\textcolor{teal}{Fact 4} & water is a kind of liquid at room temperature.  \\
\textcolor{teal}{Fact 5} & water is in the solid state , called ice , for temperatures between 0; -459; -273 and 273; 32; 0 K; F; C.  \\
Fact 6 & temperature is a measure of heat energy.  \\
\textcolor{teal}{Fact 7} & heat means heat energy.  \\
Fact 8 & water is in the liquid state , called liquid water , for temperatures between 273; 32; 0 and 373; 212; 100 K; F; C.  \\
Fact 9 & ice is a kind of solid.  \\
\textcolor{teal}{Fact 10} & if an object; a substance; a location absorbs solar energy then that object; that substance will increase in temperature.  \\
Fact 11 & melting is when solids are heated above their melting point.  \\
\textcolor{teal}{Fact 12} & adding heat means increasing temperature.  \\
Fact 13 & cooling;colder means removing;reducing;decreasing heat;temperature.  \\
Fact 14 & heating means adding heat.  \\
 \bottomrule
\end{tabular}
\caption{Example $\bm{q} \in \mathcal{D}_{qa}^{val}$ with explaining gold facts and their roles. The wrong multiple choice answer options have already been removed.
} \label{tab:pred-samples}
\end{table}

\section{Extra plots} \label{app:extra-plots}

\begin{figure}[h]
    \centering
    \begin{subfigure}{0.4\textwidth}
        \centering
        \includegraphics[width=1.0\textwidth]{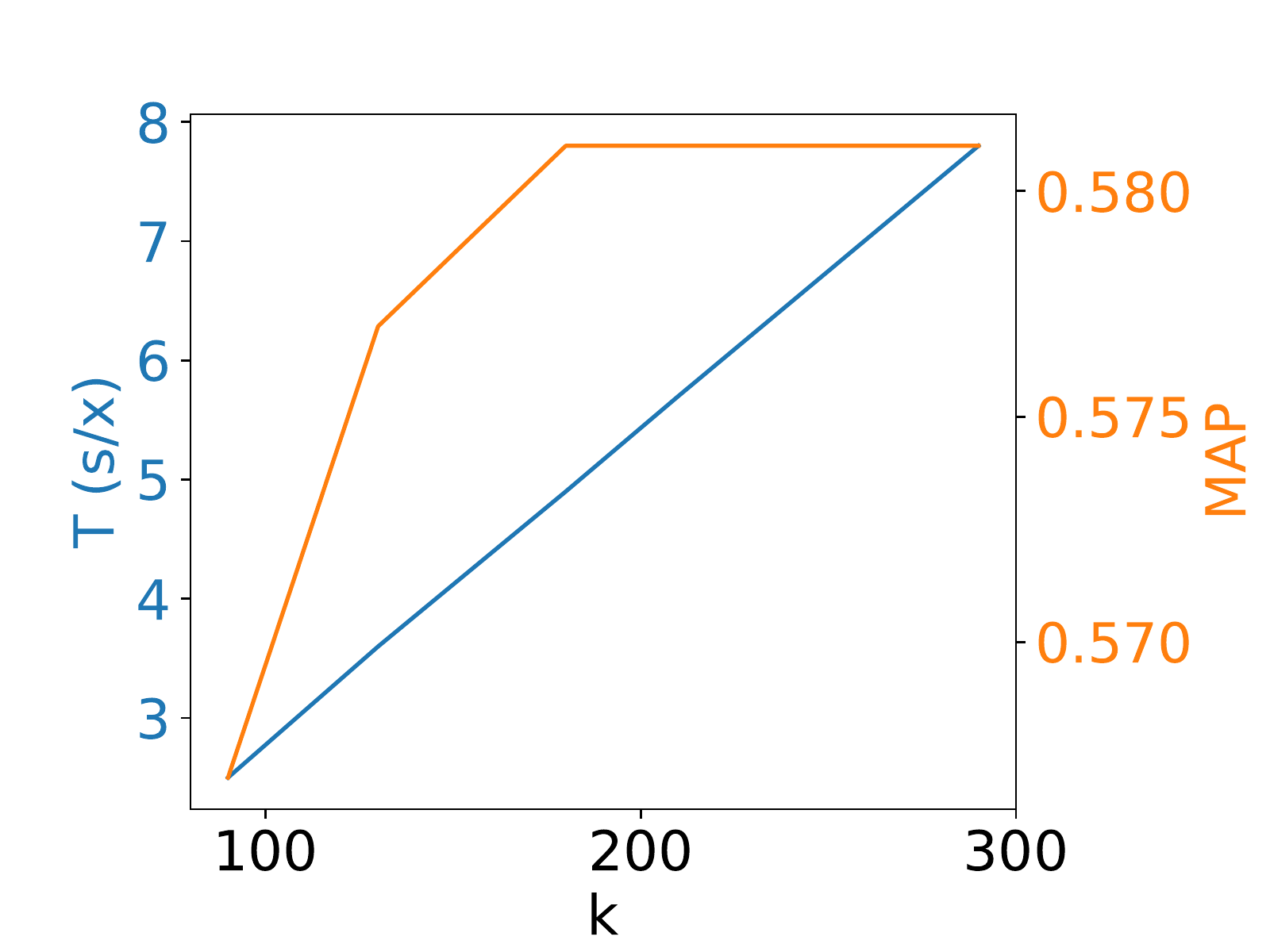}
        \caption{} 
        \label{fig:k-plot}
    \end{subfigure}
    \begin{subfigure}{0.4\textwidth}
        \centering
        \includegraphics[width=1.0\textwidth]{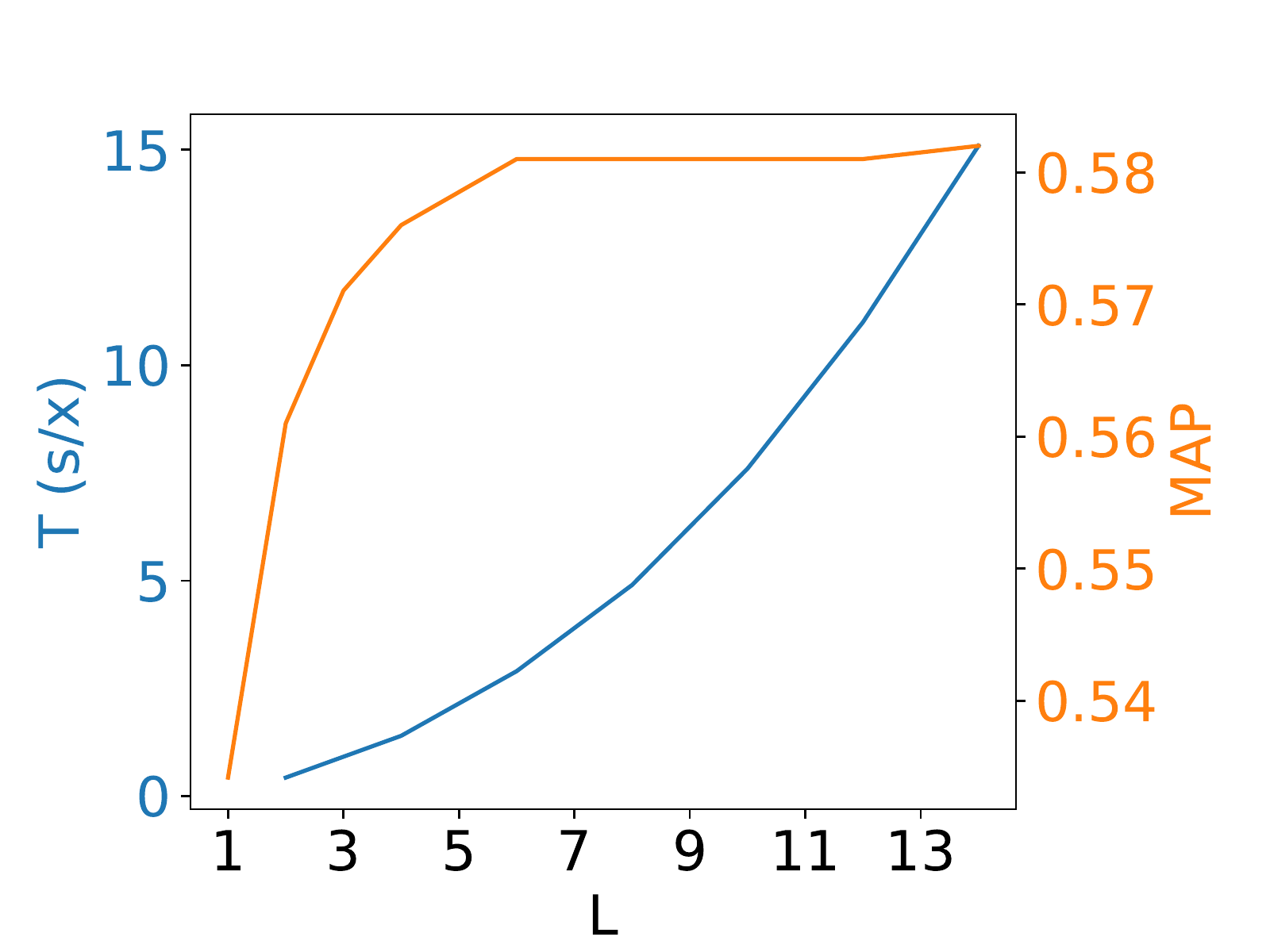}
        \caption{} 
        \label{fig:L-plot}
    \end{subfigure}
    \caption{Inference time in seconds per sample and 2020 test set MAP of ARCF for $k$ and $L$, see also tables \ref{tab:k}-\subref{tab:L}. Time grows approximately quadratically in $L$ and linearly in $k$.} 
    \label{fig:L-k-plot}
\end{figure}

See figure \ref{fig:L-k-plot}.

\end{appendices}
\end{document}